\documentclass[traditabstract]{aa} 
\usepackage{graphicx}
\usepackage{amsmath}
\usepackage{txfonts}

\begin{document}
   \title{High-impedance NbSi TES sensors for studying the cosmic microwave background radiation}

   \author{C. Nones
          \inst{1} \fnmsep\thanks{Now at the Service de Physique des Particules - CEA/DSM/IRFU/SPP - 91191 Gif-sur-Yvette - France}
          \and
          S. Marnieros \inst{1}
           \and
           A. Benoit \inst{2}
           \and
           L.Berg\'e \inst{1}
           \and
           A. Bideaud \inst{2}
           \and
           P. Camus \inst{2}
           \and
           L. Dumoulin \inst{1}
           \and
           A. Monfardini \inst{2}
           \and
          O. Rigaut  \inst{1}
          }

   \institute{CSNSM, CNRS-IN2P3,
              Bat. 108 Orsay Campus, 91400 Orsay, France\\
              \email{claudia.nones@cea.fr}
         \and
             Institut N\'eel (CNRS/UJF), 25 rue des Martyrs, 38042 Grenoble, France\\
             }


 \abstract{Precise measurements of the cosmic microwave background (CMB) are crucial in cosmology, because any proposed model of the universe must account for the features of this radiation. The CMB has a thermal blackbody spectrum at a temperature of 2.725 K, i.e. the spectrum peaks in the microwave range frequency of 160.2 GHz, corresponding to a 1.9-mm wavelength. Of all CMB measurements that the scientific community has not yet been able to perform, the CMB B-mode polarization is probably the most challenging from the instrumental point of view. The signature of primordial gravitational waves, which give rise to a B-type polarization, is one of the goals in cosmology today and amongst the first objectives in the field.\\
For this purpose, high-performance low-temperature bolometric cameras, made of thousands of pixels, are currently being developed by many groups, which will improve the sensitivity to B-mode CMB polarization by one or two orders of magnitude compared to the Planck satellite HFI detectors. \\ 
We present here a new bolometer structure that is able to increase the pixel sensitivities and to simplify the fabrication procedure. This innovative device replaces delicate membrane-based structures and eliminates the mediation of phonons: the incoming energy is directly captured and measured in the electron bath of an appropriate sensor and the thermal decoupling is achieved via the intrinsic electron-phonon decoupling of the sensor at very low temperature. \\
 Reported results come from a 204-pixel array of  Nb$_{x}$Si$_{1-x}$ transition edge sensors with a meander structure fabricated on a 2-inch silicon wafer  using electron-beam co-evaporation and a cleanroom lithography process. To validate the application of this device to CMB measurements, we have performed an optical calibration of our sample in the focal plane of a dilution cryostat test bench. We have demonstrated a light absorption close to 20\% and an optical  noise equivalent power (NEP) of about 7$\times$10$^{-16}$ W/$\sqrt{Hz}$, which is highly encouraging given the scope for improvement in this type of detectors.}

 \keywords{ cosmic microwave radiation -- instrumentation: detectors -- instrumentation: photometers -- methods: laboratory }

\titlerunning{High impedence NbSi TES to study the Cosmic Microwave Radiation}

\maketitle
%

\section{Introduction}
The energy content in the radiation coming from beyond our galaxy is dominated by the cosmic microwave background (CMB), discovered in 1965 (Penzias \& Wilson \cite{penzias}). The spectrum of the CMB is well described by a blackbody function with T = 2.725 K. This spectral form is one of the main pillars of the hot big bang model for the early Universe. 
Another observable quantity inherent in the CMB is the variation in temperature (or intensity) from one part of the microwave sky to another (Dicke et al. \cite{dicke}). Since the first detection of these anisotropies by the COBE satellite (White et al. \cite{white}), there has been intense activity to map the sky at levels of increasing sensitivity and angular resolution by ground-based and balloon-borne measurements. These were joined in 2003 by the first results from NASA's Wilkinson Microwave Anisotropy Probe (WMAP) (Hu \& Dodelson \cite{hu}), which were improved upon by analysis of the three-year and five-year WMAP data (Smoot et al. \cite{smoot}; Bennett et al. \cite{bennett}). Together, these observations have led to a stunning confirmation of the Ôstandard model of cosmology.Õ In combination with other astrophysical data, the CMB anisotropy measurements place quite precise constraints on a number of cosmological parameters, and have launched us into an era of precision cosmology. This has been confirmed by  the improved capabilities provided by the Planck satellite.

The primary science goals of CMB cosmology in the next decade are the degree-scale B-mode polarization induced by a gravitational wave background and the arcminute-scale B-mode induced by weak gravitational lensing from large-scale structure. The former will provide invaluable information on inflation and the early universe, while the latter offers a sensitive and complementary probe of the dark energy and the neutrino mass. These measurements are probably the most constraining from an instrumental point of view. To achieve these challenging goals, future instruments will require large arrays of sensitive mm-wave detectors, with wide frequency coverage for astronomical foreground monitoring, and an extensive control of polarization systematics.
Bolometers can provide photon noise-limited sensitivity in the whole CMB frequency range. The goal of many research projects, focusing on the detection of the CMB B-mode polarization, is to improve the sensitivity of the angular power spectrum measurement by at least two orders of magnitude compared to Planck (Tauber \cite{tauber}; Lamarre \cite{lamarre}).\\
 For future satellite observations, the telescope focal plane should be optimally filled with large efficient bolometer arrays, showing a noise equivalent power (NEP)sensitivity better than the CMB photon noise (NEP lower than 10$^{-18}$ W/$\sqrt{Hz}$). Several solutions are currently being developed by the different collaborations to achieve fabrication of high-performance Óbolometric camerasÓ similar to CCDs (Orlando et al. \cite{Orlando};  Essinger-Hileman et al. \cite{Essinger}; Reichborn-Kjennerud et al. \cite{Reichborn}; Battistelli et al. \cite{Battistelli}). The structure of a single pixel must therefore be compatible with a highly reliable collective fabrication and multiplexed readout electronics.


\section{Principle of operation: electron-phonon decoupling sensors}
\begin{figure}
   \centering
 \includegraphics[width=6 cm]{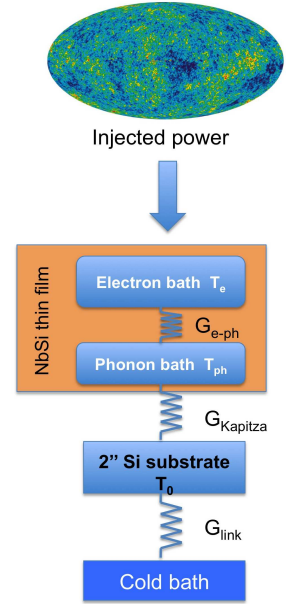}
      \caption{(Color online) Schematic view of the thermal model.}
         \label{FigThermalModel}
   \end{figure}

We focus here on bolometers that measure infrared radiation in the mm range by the resulting temperature fluctuations. The basic structure of a single pixel consists of three important elements: a radiation absorber, a thermometer, and a thermally isolated holder enabling a very weak coupling of the former two elements with respect to the cryostat cold bath. The various existing detectors differ in the technique used for absorbing the incident light (horns, absorbing films, antennas), the nature of the thermometer (high-impedance type, or superconducting transition edge sensors, TES) and in the solution given to the thermal decoupling problem (diversely manufactured membranes, micromesh membranes). The latter is an extremely critical point. Indeed, the sensitivity of a bolometer to an incident power P is limited by its noise equivalent power: NEP$^2$ = 4k$_B$T$^2$G, and therefore directly depends on the thermal decoupling G between the absorber and the thermometer on one hand and the cold bath on the other. In most of the cases this thermal decoupling is provided by micromeshed membranes that require very delicate cleanroom processes.
We present here an innovative device that replaces these delicate membrane-based structures and eliminates the mediation of phonons for the energy transfer: the incoming energy is directly captured and measured in the electron bath of an appropriate TES sensor and the thermal decoupling is achieved via the natural decoupling that exists between electrons and phonons of the sensor at very low temperature (see Fig. \ref{FigThermalModel}).
Electron-phonon decoupling to replace membranes has already been proposed for TES sensors that were coupled to radiation by antennas (Karasik et al. \cite{Karasik}). We developed direct light absorption into amorphous Nb$_{x}$Si$_{1-x}$ thin-film sensors, known to undergo a superconductor-metal-insulator transition upon the niobium concentration and thickness (Marrache et al. \cite{marrache}). 

To be suitable for array fabrication and light absorption, the sensor must have a size of the order of the wavelength ($\sim$ 1 mm) and meet the vacuum impedance.\footnote{We recall that the optimal absorption of an incident electromagnetic wave in a quarter wave cavity configuration is obtained for $R_{\Box sensor} \sim Z_0 = \frac {E}{H}=377  \Omega$ per square into the vacuum (Hadley et al. \cite{hadley}; Bauer et al. \cite{bauer}). In our case the calibration was realized without a reflector leading to a light absorption that is not optimal. For the ultimate setup the detector array will be processed on a $\lambda$/4 thick Si wafer, with a metallic reflecting foil on its backside.} 
The detected incoming photons will see the normal resistance of our TES because their energies are higher than the superconducting gap: $h\nu >> k_BT_c$. We have previously shown (Marnieros \cite{marnieros}; Marrache et al. \cite{marrache}) that Nb$_{x}$Si$_{1-x}$ is a compound where the superconducting temperature T$_c$ and the normal resistance R$_N$ can be separately tuned through the Nb concentration ''x" and the film thickness. This device could be combined with an optimal readout system. Keeping a constant normal impedance with respect to the incoming electromagnetic wave, the TES transition resistance could be adapted to both a SQUID-based readout via interleaved electrodes (R$<$1$\Omega$)  or to JFET transistor readout via a meander shape of the film (R$>$1M$\Omega$) (Marnieros et al. \cite{marnieros2}). In this paper we will discuss the latter case. Interestingly the Time Domain Multiplexing technique (TDM) has already been developed for such high-impedance devices (Benoit et al. \cite{benoit_TDM}).

\section{NbSi TES sensors with a meander structure}
\subsection{Sample preparation and fabrication process}
\begin{figure}
   \centering
 \includegraphics[width=8 cm]{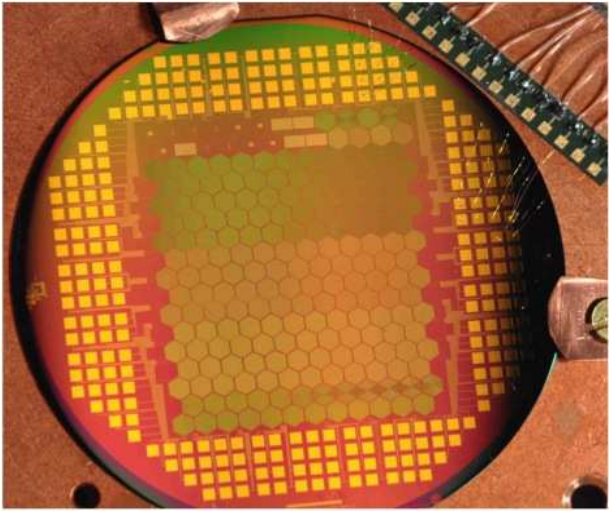}
      \caption{(Color online) Picture of our 204-pixel device.}
         \label{FigMatrix}
   \end{figure}

\begin{figure}
   \centering
 \includegraphics[width=8 cm]{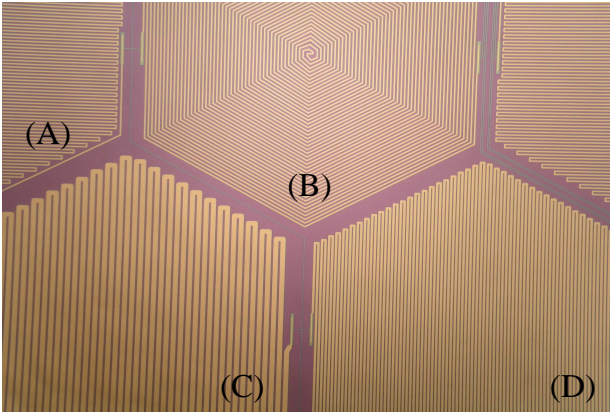}
      \caption{(Color online) Details of the meander structure of pixels. The different geometries described in Table \ref{table:geometries} are visible: A - thin horizontal; B - thin circular; C - thick vertical; D - thin vertical.}
         \label{FigMeander}
   \end{figure}

Our sample is a 204-pixel array (Fig. \ref{FigMatrix}, Fig. \ref{FigMeander}) composed of amorphous Nb$_{x}$Si$_{1-x}$ TES with a meander structure, niobium leads and gold contact pads. We present here results obtained with a 50-nm thick Nb$_{0.14}$Si$_{0.86}$ TES sample. 
Different geometries, summarized in Table \ref{table:geometries}, were produced on the same wafer to study the absorption sensitivity to different light polarizations as described below.  In particular, we changed the orientation of the meander structure (vertical, horizontal or circular), the width of the NbSi lines and the gap between two NbSi lines. 

\begin{table}
\caption{Pixel geometries: different geometries were produced to study the absorption sensitivity to different light polarizations; d is the width of NbSi meander lines  while the gap is the distance between two Nb lines.}         
\label{table:geometries}     
\centering                          
\begin{tabular}{c c c c}        
\hline\hline                
Geometry name & d [$\mu m$] & gap [$\mu m$] & surface  [$mm^2$]\\    
\hline                        
   Thin horizontal (A)$^{\mathrm{*}}$& 10 & 10 & 1.51 \\
   Thin circular (B)& 10 & 10    &  1.55\\
   Thick vertical (C)& 30 & 10     &  2.25\\
   Thin vertical (D)& 10 & 10 & 1.54 \\
   Medium vertical & 15 & 5    &  2.31\\
\hline                                  
\end{tabular}
\begin{list}{}{}
\item[$^{\mathrm{*}}$] Capital letters refer to Fig. \ref{FigMeander}.
\end{list}
\end{table}

The steps of our simplified but highly reliable micro-fabrication procedure are described below:
\begin{enumerate}
\item 1-$\mu$m thick SiN deposition by LPCVD (low pressure chemical vapor deposition) on a two-inch silicon wafer. This layer is for electrical insulation between the TES and the Si wafer for testing purposes at room temperature.
\item Co-evaporation of a layer of  Nb$_{x}$Si$_{1-x}$ (x=0.14, 50 nm thick): NbSi thin film is manufactured by electron-beam co-evaporation of Nb and Si. 
\item Evaporation of a 25-nm thick Nb layer.
\item Reactive ion etching (RIE)  technology is then used to produce the meander structure of pixels (NbSi) and the electrical contacts (NbSi + Nb). 
\item Au evaporation and lift-off: a gold layer (100-150 nm) is deposited on the wafer to form the square electrical contact pads.
\end{enumerate}

\subsection{Electronic properties of the NbSi TES}
\label{sec:thermal_prop}
\begin{figure}
   \centering
 \includegraphics[width=9 cm]{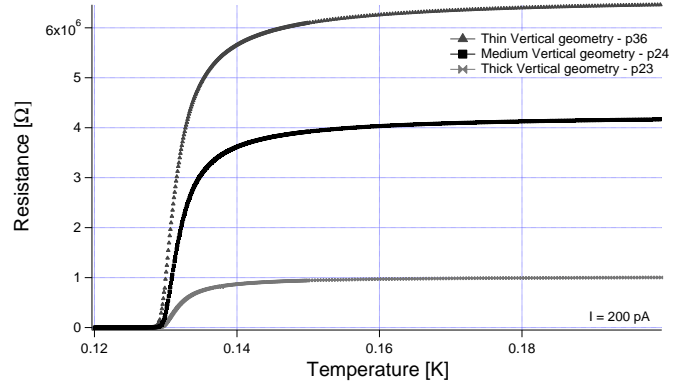}
      \caption{R vs. T plot for three different pixel geometries for a low current excitation.}
         \label{FigR_T}
   \end{figure}

The device was first cooled down in a dilution refrigerator at CSNSM-Orsay to test its electronic properties, such as the normal state resistance, the critical temperature, the homogeneity between pixels, and the electron-phonon decoupling. Electron-phonon decoupling is a useful parameter to evaluate the light absorption coefficient, as  will be described below.\\
Figure \ref{FigR_T} shows the resistance versus temperature plot for pixels with three different geometries under low current bias (200 pA). Their normal resistance scales accordingly with the pixel geometry (1 M$\Omega$ to 6 M$\Omega$ upon the meander length/width ratio). All pixels exhibit a critical temperature transition close to 135 mK with a transition width of about 10 mK.

\begin{figure}
   \centering
 \includegraphics[width=9 cm]{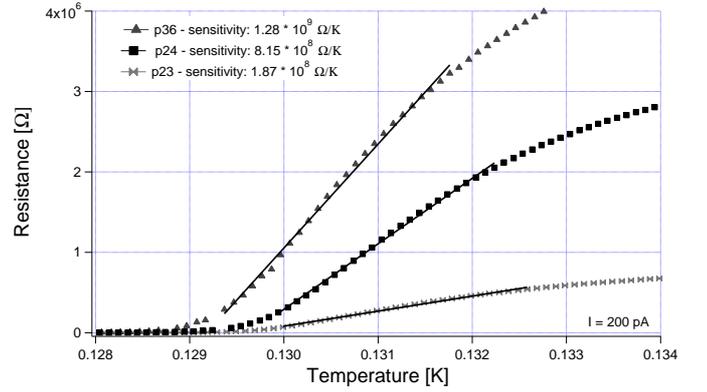}
      \caption{Zoom of Fig. \ref{FigR_T} to better show the temperature sensitivity.}
         \label{FigSensitivity}
   \end{figure}

An important parameter for this kind of device is the temperature sensitivity, obtained by making a linear fit of  the R vs. T curves as shown in Fig. \ref{FigSensitivity}.  The dimensionless sensitivity dlog(R)/dlog(T) for all pixels is close to 100 at the lower part of the transition, which is a typical value for TES. The particular meander shape of each sensor apparently does not affect the sensitivity. 

\begin{figure}
   \centering
 \includegraphics[width=9 cm]{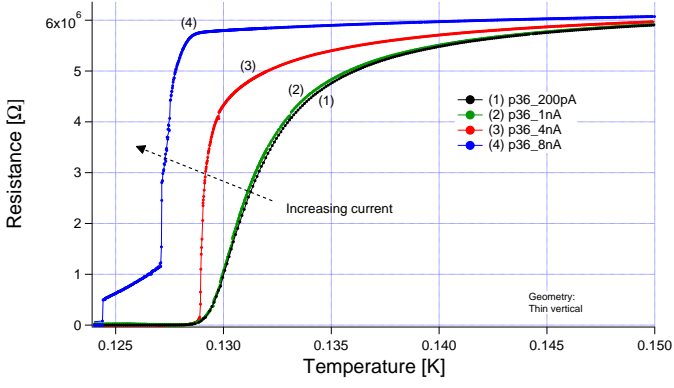}
      \caption{(Color online) R vs. T plot for different polarization currents for the thin geometry. Because of the positive electro-thermal feedback, high current bias induces instabilities on the R vs. T curves (e.g. 8-nA curve).}
         \label{FigR_T_polar}
   \end{figure}

We now consider the case of the thin meander geometry (10-$\mu m$  wide NbSi meander lines with 10-$\mu m$  gap between the lines). When performing R vs. T measurements for the same pixel, we notice that the curves slightly shift to lower temperature values as the bias current increases (Fig. \ref{FigR_T_polar}).

Our data fit the electron-phonon coupling model very well,  a model that is well established for metals but frequently used also for superconductors with low temperature transitions (Hart et al. \cite{Hart}; Hoevers et al. \cite{Hoevers}; Karasik et al. \cite{Karasik2}). This model, sketched in Fig. \ref{FigThermalModel},  defines an electron bath and a phonon bath that are thermally coupled via the so-called electron-phonon thermal conductance, $G_{e-ph}$. When a bias current is applied, the electron temperature rises above the phonon temperature due to the Joule heating effect. The phonon bath of the sensor is coupled to the Si substrate via the Kapitza thermal conductance, $G_{Kapitza}$. According to this model, the relation between the electron temperature and the phonon temperature for a given electrical power ($P$) is described by the following formula:

\begin{equation}
      \frac{P}{\Omega}=g_{e-ph} \cdot (T_e^\beta-T_{ph}^\beta) ,
   \end{equation}

where $T_e$ and $T_{ph}$  are the electron and phonon temperatures, respectively, $\Omega$ the volume of NbSi TES thin film, $g_{e-ph}$  the electron-phonon coupling constant, and $\beta$=5 in the case of usual metals (Cu, Ag, Au ...) (Roukes et al. \cite{roukes}; Wellstood et al. \cite{wellstood}). 
The thermal conductance between electrons and phonons is then given by

\begin{equation}
      G=(\partial P / \partial T_e)_{T_{ph}} = g_{e-ph} \cdot \beta \cdot T_e^{\beta-1} \cdot  \Omega .
      \label{EqG}
   \end{equation}

In our case the phonon temperature of the NbSi TES is taken equal to the substrate temperature ($T_{ph}  \sim T_0$) because of the high Kapitza thermal conductivity, . 
To extract $T_e$, we assume that the resistance of our TES depends only on the temperature of the electron bath ($R=R(T_e)$). At high bias, $T_e$ increases due to the Joule heating of the electrons and $R(T_e)$ is rising, up to the normal state resistance ($R_N$). We measure $g_{e-ph} \sim 200 \; W/K^5cm^3$, and $G_{e-ph} \sim 2.2 \cdot 10^{-8} \;  W/K$. This natural decoupling between the electrodes and the phonons of the NbSi will replace the delicate membrane-based decoupling mechanism used in standard bolometric arrays. The resulting thermal conductance is quite high compared to that of membranes owing to the large NbSi volume. In the present case, the NbSi size was adjusted to optimize light absorption. Nevertheless, some alternative solutions with substantially lower thermal conductance are discussed in Section \ref{Conclusion}. \\
The time constant of our device has been estimated through the response to cosmic ray interactions. The observed pulses show a decay-time constant of 50 $\mu s$ at 130 mK.

\section{Light absorption measurements}

\begin{figure}
   \centering
 \includegraphics[width=9 cm]{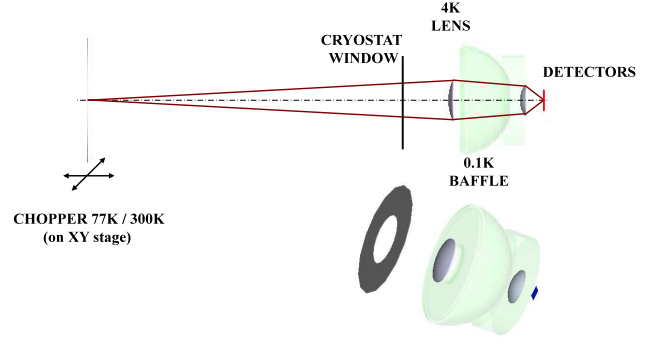}
      \caption{(Color online) 3D scheme of the optical design. }
         \label{Fig_optical}
   \end{figure}

To validate this device for astrophysical applications, it was necessary to measure its capability to absorb light. To this purpose the sample holder was mounted in the focal plane of a custom-designed dilution cryostat with optical access. The optical path was equipped with a 125-170 GHz bandpass filter stack. \\
The cooling power of the cryostat is $\sim$ 100 $\mu W$ at 100 mK and its operation, including the gas handling system, is automated and controlled remotely (Monfardini et al. \cite{monfa1}). 
The optical design is displayed in Fig. \ref{Fig_optical}. The telescope focal plane is re-imaged with a bi-polynomial field mirror and two high-density polyethylene (HDPE, n=1.56) lenses. To reduce the aberrations introduced by the field mirror, the deflection angle is kept as low as possible, the lower limit being determined by the cryostat physical dimensions. The de-magnifying factor is $\sim$6, giving an effective aperture on the detector plane of f/1.6. In its present configuration, the instrument admits up to a 100-mm diameter focal plane with an aperture as low as f/1.6. The re-imaging optics are telecentric in image space, with the chief rays parallel to the optical axis. Thus, the illumination on the focal plane is everywhere perpendicular to the detectors with the light-cones subtending the same angle at every point. This homogeneity eliminates any position-dependence across the array arising from the overlap of the fixed angular-radiation pattern of the pixels and the incident radiation.

To eliminate any spurious off-axis radiation, which can significantly degrade the detector performance, a novel cold baffling system was designed and fabricated. It is based on a three-stage stack of properly shaped aluminum reflecting surfaces. Each stage consists of a cross-section of an ellipsoid shaped in such a way that any stray rays entering the optical system are back-reflected. In addition to this baffling system, infrared-blocking and band-defining filters were installed on the radiation screens at 80, 5, and 1 K and 100 mK. These consist of multi-layer cross-mesh filters, each augmented with a layer of Zitex$^{\textregistered}$ film. The pass band of this filter stack is 125-170 GHz.

\subsection{Chopper measurement}
\begin{figure}
   \centering
 \includegraphics[width=9 cm]{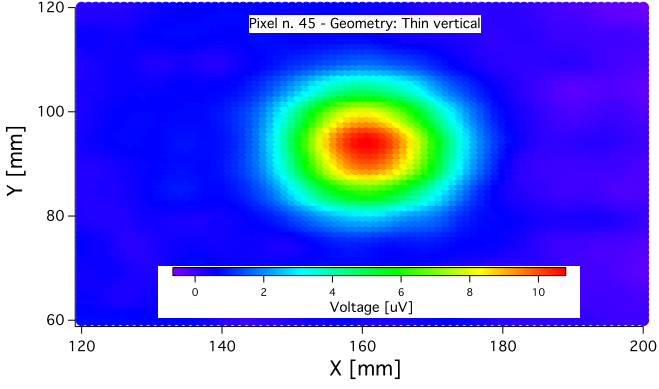}
      \caption{(Color online) Color palette plot for one pixel showing a signal of 10.8 $\mu$V in the light absorption generated by a blackbody radiation whose temperature is modulated between 77 and 300 K. X and Y axis of the plot correspond to the position - in mm - of the blackbody hole in the front focal plane, outside the cryostat. }
         \label{FigCarta}
   \end{figure}

To calibrate the light absorption efficiency of our detector, we used a setup based on a blackbody radiation chopped between 77 K and 300 K. The focal plane outside the cryostat was filled with Ecosorb$^{\textregistered}$ immersed in liquid nitrogen, simulating a 77 K blackbody. A small 1-cm hole through the Ecosorb$^{\textregistered}$ was made, behind which a mechanical chopper (turning wheel) could alternate between 77 K and 300 K (Fig.\ref{Fig_optical}). The whole setup could be precisely shifted along the X-Y direction of the focal plane, allowing a raster scan of the image across the detection array. The details of this measurement system have been previosly discussed elsewhere (Benoit et al. \cite{benoit}; Swenson et al. \cite{swenson}). We performed a complete scan of the array, showing an average TES response of 10-12 $\mu$V to the chopper modulated light for pixels with thin vertical geometry. We point out that, due to a problem in the design of the array holder, pixels at the boundary of the array are not homogeneously illuminated and are consequently characterized by a reduced signal. We will not take them into account for the calibration of the absorption efficiency.  Fig. \ref{FigCarta}  shows a typical result for a central pixel that is not affected by boundary problems. We observed negligible crosstalk when the modulated image focused outside the pixel. This effect clearly demonstrates our scheme of direct absorption of light by the electron bath of the NbSi (Fig. \ref{FigThermalModel}). The Si substrate is not affected by light absorption, in contrast to the illuminated NbSi sensors whose temperature responds to the blackbody modulation. A quantitative analysis of the light absorption coefficient is given in Section \ref{SecResults}.

\subsection{Polarizer measurement}

\begin{figure}
   \centering
 \includegraphics[width=9 cm]{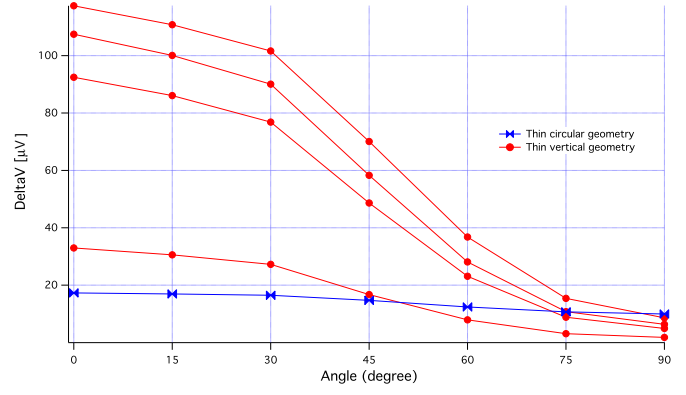}
      \caption{(Color online) $\Delta V$ as a function of the rotation angle of the polarizer. Pixels respond to polarized light according to their geometry.}
         \label{FigPolarizer}
   \end{figure}

We also calibrated of our sample to polarized radiation to study the impact of the various meander sensor geometries to light absorption. A linear polarizer was positioned in front of the cryostat optics and rotated in 15$^{\circ}$ steps. The detector response to polarized light modulated between 77 K and 300 K is given in Fig. \ref{FigPolarizer}. The behavior of the output signal with respect to the polarization angle depends on the pixel geometry. Incoming light polarized parallel to the meander lines of a given sensor is absorbed with a much higher efficiency. For the quasi-circular meander pixels (geometry B of Fig. \ref{FigMeander}) we observe a very flat response. There is considerable dispersion in the amplitude of the signals for the "vertical geometry" that is correlated to the distance of the pixels from the center of the array. As mentioned before, toward the edge of the array the detector holder is partially screening the focal plane. The lower two curves come from pixels located on the first line of the array and are clearly affected by this effect. We can define the sensitivity of our device to polarization by calculating the "contrast" $\Delta V_{max}-\Delta V_{min}/\Delta V_{max}+\Delta V_{min}$. The polarization selectivity of the linear shape meanders is very high, their contrast being close to 90\% (even for the side pixels). This is not the case for the quasi-circular geometry pixels that exhibit a contrast of approximately 10\%. This effect must be taken into consideration for the light absorption calibrations and more generally for the design of NbSi arrays dedicated to astrophysical observations.

\section{Light absorption results}
\label{SecResults} 
To evaluate the light absorption efficiency of our device, we combined the results obtained from the study of the NbSi TES electronic properties with the measurements of its response to a modulated infrared radiation source.
The temperature modulation $\Delta T$ of an illuminated TES pixel is given by the following relationship:

\begin{equation}
      \Delta T=\epsilon \cdot \frac{\Delta P}{G_{e-ph}},
 \end{equation}
 where $\epsilon$ is the light absorption coefficient, $\Delta P$ the light power incident on the pixel, and G$_{e-ph}$ the electron-phonon thermal conductance of the NbSi TES.

 The absorption coefficient $\epsilon$ is deduced by independently measuring G$_{e-ph}$, $\Delta T$ and $\Delta P$:
 
 \begin{itemize}
\item[$\bullet$] {\bf G$_{e-ph}$}: as already mentioned, NbSi sensors are well described by an electron-phonon decoupling model, following the law given by Eq. \ref{EqG}. In this model, the electron-phonon thermal conductance G$_{e-ph}$ strongly depends on the electron temperature:
 
 \begin{equation}
      G_{e-ph}= \Omega \cdot g_{e-ph} \cdot 5 \cdot T_{e}^4 .
   \end{equation}
   
 The precise value of G$_{e-ph}$ has been given in Section \ref{sec:thermal_prop}.

 \item[$\bullet$] {\bf $\Delta P$}: It is possible to estimate the power on the pixels by considering the radiation spectrum emitted by the modulated blackbody source and the total transmission of the cryostat optics (filter + lenses + cold pupil). We deduced a power modulation of  1.3 pW per pixel.

 \item[$\bullet$] {\bf $\Delta T$}: temperature fluctuations are directly related to the resistance of our TES via the sensitivity $\alpha=\Delta R/\Delta T$. For a constant-current biased TES we have $\Delta T= \Delta T/\Delta R \cdot \Delta V/I=1/\alpha \cdot \Delta V/I$.
 
\end{itemize}

 Taking into account all these quantities, it is straightforward to derive the following expression for the light absorption coefficient value:
 \begin{equation}
      \epsilon=\frac{\Delta V}{I} \cdot  \frac{1}{\alpha} \cdot g_{e-ph} \cdot 5 \cdot T_{e}^4 \cdot \frac{\Omega}{\Delta P} .
   \end{equation}

\begin{figure}
   \centering
 \includegraphics[width=9 cm]{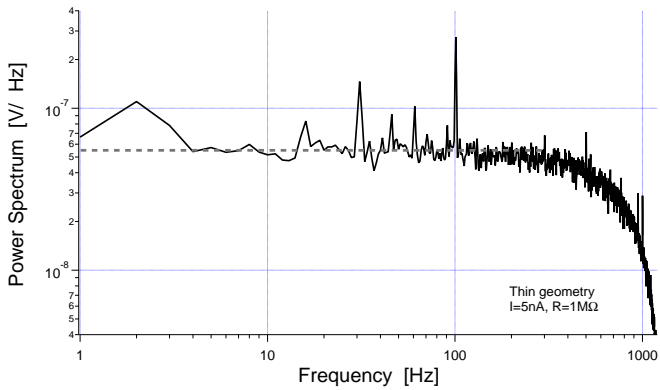}
      \caption{Noise spectrum of a thin geometry pixel acquired with a lock-in amplifier at 1250 Hz. The dotted line shows the average noise between 10-100 Hz.}
         \label{FigNoise}
   \end{figure}

Given the specific meander geometry of our TES,  $\epsilon$ will depend on the polarization of the incident light. We obtain  $\epsilon = 20\%$ for light with polarization parallel to the meanders and 10\% for non-polarized light. Our TESs are indeed transparent to photons polarized perpendicularly to the meander lines. The ultimate performances measured in our setup, including the 0.2 quantum efficiency (parallel polarized light), can be given in terms of NEP:
$NEP=7 \cdot 10^{-16} \;  W/\sqrt Hz$. The noise figure is flat down to a few Hz. Below this value our TES were limited by residual temperature fluctuations of the cryostat.  Fig. \ref{FigNoise} shows the noise spectrum acquired with a lock-in amplifier at 1250 Hz.

\section{Conclusions and perspectives}
\label{Conclusion}
We have presented low-temperature data showing the direct absorption of infrared radiation by NbSi TES thermometers  with an efficiency of 20\%. Electron-phonon decoupling inside the NbSi film was used to replace the micromesh membranes, which are rather fragile and too delicate to be produced for large arrays. The measured performance of our detector is very promising and substantial further progress can be easily achieved. Indeed, we can optimize the optical absorption coefficient and electron-phonon decoupling as described below.
The first important improvement on the optical design of the detector is to introduce a $\lambda$/4 reflector. In the present setup Ecosorb$^{\textregistered}$ foam absorber was placed on the rear of the Si wafer to stop light from passing through the NbSi array. The absorption coefficient on our future detectors can be improved using $\lambda$/4-thick Si substrates, $\lambda$ being the wavelength of the incident radiation. The NbSi TES will be placed on the front side of the Si wafer and a reflecting film will be deposited on the back side. The thickness of the NbSi TES also has a direct impact on the absorption coefficient and will be fine-tuned in the future.
Thermal coupling $G$ is the second parameter to deal with to reduce the bolometer NEP noise (NEP =$\sqrt(4K_BT^2G$)).
The particularity of the electron-phonon coupling presented here is its high dependence on temperature (G$_{e-ph}\sim T^{4}$). Operation of our detector at around 60 mK rather than 130 mK will reduce the NEP by a factor of ten. At 50 mK refrigerators, equipped with an optical window coupled to a telescope, are developed for a variety of applications including future satellites. Low-temperature operation of such detectors is therefore no longer a problem. These improvements should bring the NEP of our device close to $10^{-17}$ W/$\sqrt{Hz}$, which is sufficient for the majority of ground-based applications.
Another solution to reduce G$_{e-ph}$ is reduce the TES size. Indeed, electron-phonon coupling is proportional to the NbSi volume. It will be interesting to test an array with absorption of light by antennas, dissipating their energy directly to an NbSi TES film with a very small surface. In this case the size of the NbSi can be substantially smaller than the radiation wavelength and should only be adjusted to the antenna's impedance. Reducing both its thickness and surface,  we can reasonably operate with a 5000 smaller NbSi sensor. The resulting NEP will therefore be close to  $10^{-18}$ W/$\sqrt{Hz}$. We can also consider a setup without antennas but with an array of microlenses in front of our array, similar to that proposed by SRON for their kinetic inductance detectors (Monfardini et al. \cite{monfa2}). In this case the incident radiation is focused on a much smaller area of each pixel and permits a downscale of NbSi sensor dimensions. In this configuration, the reduction of the sensor surface is typically a factor 100, implying an NEP $<  10^{-17}$ W/$\sqrt{Hz}$.
Given the very promising results described and discussed above, we can plan future strategies for the improvement of this device, depending on specific Earth-, balloon- or space -based applications.

\begin{acknowledgements}
      Part of this work was supported by  the French ANR agency (project  "ANR-06-BLAN-0326), the CNES agency (project "DCMB" and "BSD") and by a Marie Curie Intra European Fellowship within the 7th European Community Framework Programme FP7/2007-2013 (Proposal No. 236122).
\end{acknowledgements}

\end{document}